\documentclass{iaus}
\usepackage{natbib}
\usepackage{url}
\usepackage{graphicx}

\title[Can massive PopIII avoid PISN ?] 
{Can very massive stars avoid Pair-Instability Supernovae?}

\author[Ekstr\"om \& al.]   
{Sylvia Ekstr\"om,
 Georges Meynet
 \and Andr\'e Maeder}

\affiliation{Geneva Observatory, University of Geneva, Maillettes 51 - CH 1290 Sauverny, Switzerland}

\pubyear{2008}
\volume{IAUS250}  
\jname{Massive Stars as Cosmic Engines}
\editors{F. Bresolin, P.A. Crowther \& J. Puls., eds.}

\begin{document}

\maketitle

\begin{abstract}
Very massive primordial stars ($140\ M_{\odot} < M < 260\ M_{\odot}$) are supposed to end their lives as PISN. Such an event can be traced by a typical chemical signature in low metallicity stars, but at the present time, this signature is lacking in the extremely metal-poor stars we are able to observe. Does it mean that those very massive objects were not formed, contrarily to the primordial star formation scenarios ? Could it be possible that they avoided this tragical fate ?

We explore the effects of rotation, anisotropical mass loss and magnetic field on the core size of very massive Population III models. We find that magnetic fields provide the strong coupling that is lacking in standard evolution metal-free models and our $150\ M_{\odot}$ Population III model avoids indeed the pair-instability explosion.
\end{abstract}

\firstsection 
\section{Introduction}

According to \citet{HFWLH03}, the fate of single stars depends on their He-core mass ($M_{\alpha}$) at the end of the evolution. They have shown that at very low metallicity, the stars having $64\ M_{\odot} < M_{\alpha} < 133\ M_{\odot}$ will undergo pair-instability and be entirely disrupted by the subsequent supernova. This mass range in $M_{\alpha}$ has been related to the initial mass the star must have on the main sequence (MS) through standard evolution models: $140\ M_{\odot} < M_{\rm ini} < 260\ M_{\odot}$. Since we will present here a non-standard evolution, we will rather keep in mind the $M_{\alpha}$ range.

The typical mass of Population III (Pop III) stars is explored by early structure formation studies and chemistry considerations about cooling. Different studies \citep[see][among others]{ABN02,BCL02} give the same conclusion: Pop III stars are supposed to be massive or very massive, even when a bimodal mass distribution allows the formation of lower mass components \citep[see][]{NakUm01}. Therefore we expect that many among them should die as pair-instability supernovae (PISN).

\subsection{A typical chemical signature which remains unobserved}

These PISN events are supposed to leave a typical chemical signature. According to \citet{HW02}, the complete disruption of the star leads to a very strong odd-even effect: the absence of stable post-He burning stages deprives the star of the neutron excess needed to produce significant amounts of odd-$Z$ nuclei. Also the lack of $r$- and $s$-process stops the nucleosynthesis around zinc. Even if one mixes these yields with the yields of zero-metallicity $12-40\ M_{\odot}$\ models (which end up as standard Type II SNe), the PISN signature remains and should be observable.

However, using those yields, \citet{TVS04} have shown that it provides only a very poor fit to the abundances pattern observed in the metal-poor stars known today. The odd-even effect is not observed, and the models significantly over-produce Cr and under-produce V, Co and Zn.

The most metal-deficient stars are supposed to be formed in a medium enriched by only one or a few SNe. The absence of the chemical signature of the PISN is a strong argument against their existence. But how could that be?

\subsection{Simple solutions}

The simplest solution to explain this absence is to suppose that the mass domain in question was not formed in the primordial clouds. Maybe the primordial IMF was not as top-heavy as we actually think, and the most massive stars formed then could very well be too small for such a fate. However, recent works on primordial stars formation seem to rule out this possibility \citep[see][]{yoshomha06,oshnorm07}.

Another possibility is that the signature was very quickly erased by the next generations of stars. Maybe the metal-poor stars we observe are enriched by more SNe than we actually think, and the later contributions are masking the primordial ones. Only the observation of more and more metal-deficient stars will provide an answer to that possibility.

One can also wonder whether there would be a way for those stars to avoid their fate. In this context, the simplest solution is to suppose that some mechanism could lead to such a high mass loss that the conditions for pair-instability would no more be met in the central regions. The aim of the present work is to explore this possibility.
\section{Rotation at very low metallicity}

Radiatively-driven winds are supposed to scale with metallicity as $Z^{(0.5-0.86)}$. Thus, very low or even no mass loss is expected when Z approaches to 0.

\begin{figure}[b]
\begin{center}
 \includegraphics[width=.3\textwidth]{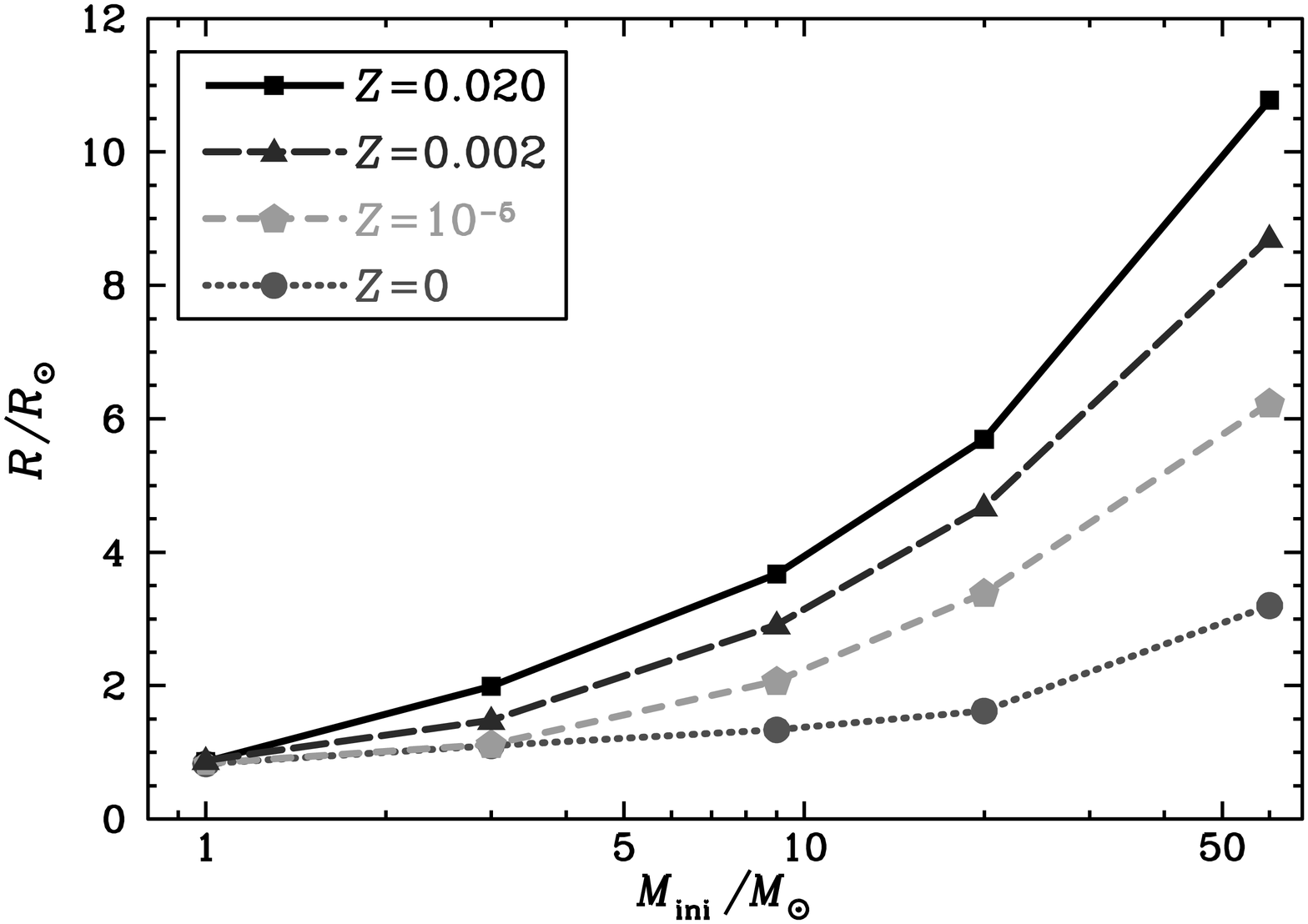}\hspace{.015\textwidth}
 \includegraphics[width=.3\textwidth]{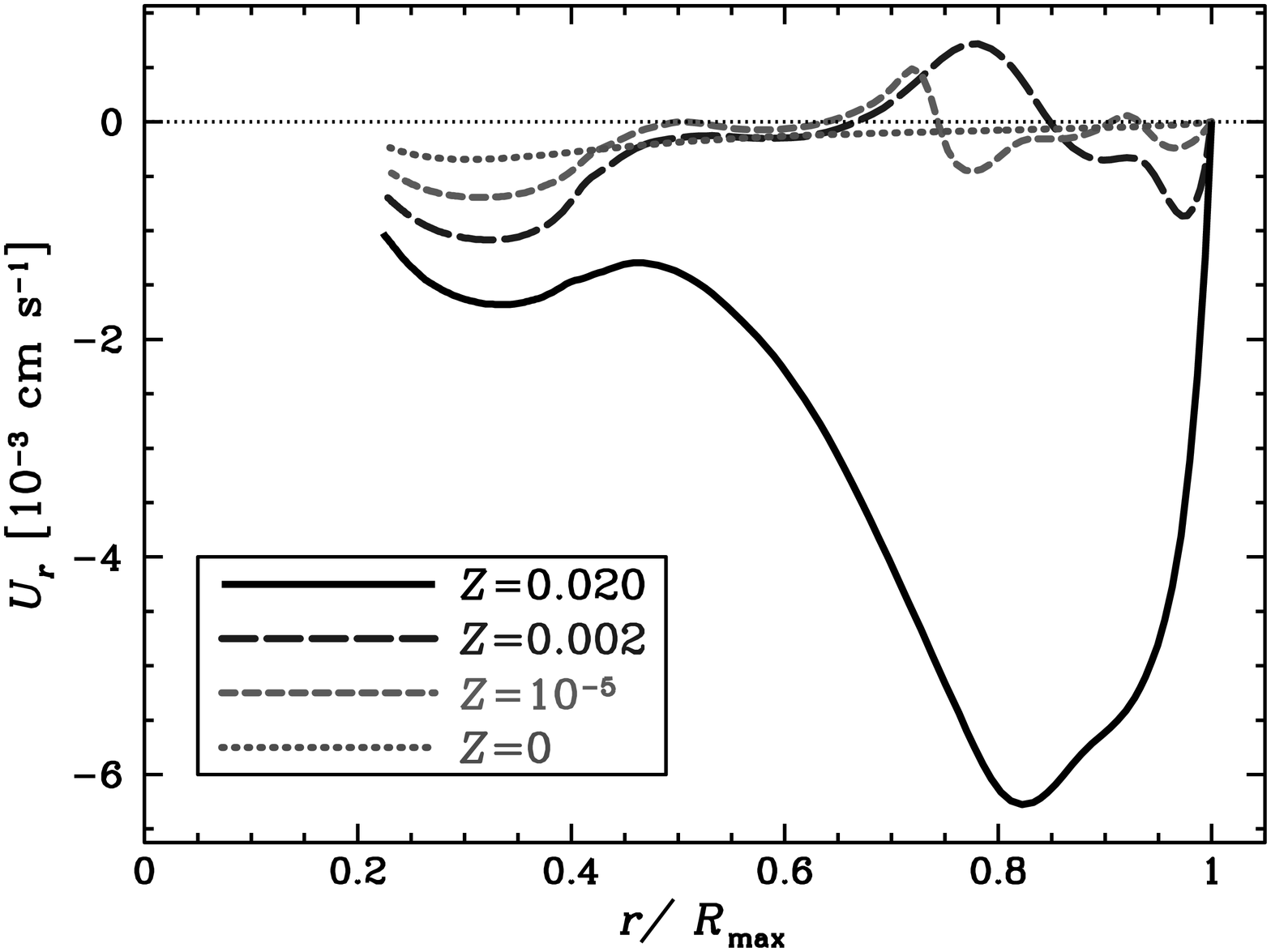}\hspace{.015\textwidth}
 \includegraphics[width=.3\textwidth]{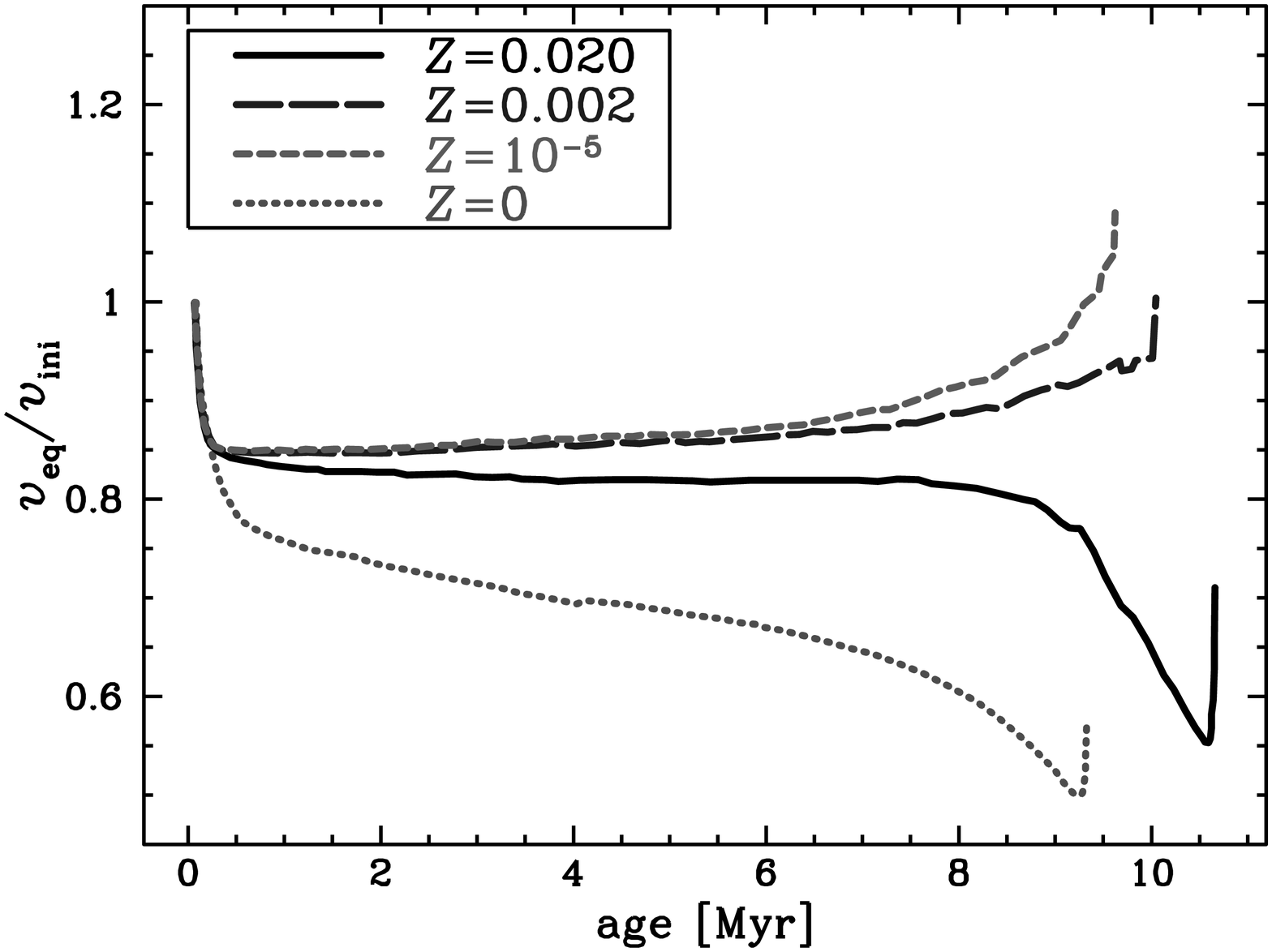}
 \caption{\emph{Left:} Variations in the radius as a function of initial mass for various metallicities. \emph{Centre:} Amplitude of the radial component of the meridional circulation inside $20\ M_{\odot}$ models at the same evolutionary stage ($X_\mathrm{c}=0.4$) and having the same rotation rate ($\upsilon/\upsilon_\mathrm{crit}=0.5$) but various metallicities. \emph{Right:} Evolution of the equatorial velocity during the MS. The models are taken from \citet{emmb08}.}
   \label{frurv}
\end{center}
\end{figure}

However, we have shown \citep{MEM06} that rotation can change the mass loss history of low metallicity stars in a dramatic way. Two processes are involved:
\begin{enumerate}
\item During MS, because of the low radiative winds, the star loses very little mass and thus very little angular momentum. As the evolution proceeds, the stellar core contracts and spins up. If a coupling exists between the core and the envelope (\textit{i.e.} meridional currents or magnetic fields), the surface may be accelerated up to the critical velocity and the star may lose mass by a mechanical wind due to centrifugal acceleration. The matter is launched into a decretion disk, which may be dissipated later by the radiation field of the star.
\item Rotation induces an internal mixing that enriches the surface in heavy elements. The effective surface metallicity is enhanced by a factor that has been shown to be very large (up to $10^6$ for a $60\ M_{\odot}$ at $Z_{\rm ini}=10^{-8}$). Rotation also favours a redward evolution after the MS, allowing the star to spend more time in the cooler part of the HR diagram. The opacity of the envelope is increased, and the radiative winds may thus be drastically enhanced.
\end{enumerate}

Both effects add up and lead to strong mass loss at very low metallicity.

\subsection{Rotation at $Z=0$}

But what happens when the metallicity drops down to $Z=0$ strictly? The absence of carbon prevents the star to start burning hydrogen through CNO-cycle, but $pp-$chains are not energetic enough to sustain the star, so it contracts longer during its formation. At metallicities $Z=0.020,\ 0.002,\ 10^{-5}$, and 0 respectively, the radii on the ZAMS will be in a ratio 3.5~:~2.8~:~2~:~1 for a $20\ M_{\odot}$ model (see Fig.~\ref{frurv}, \emph{left}), that is the $Z=0$ star will be twice as compact as the $Z=10^{-5}$ one. This has a direct influence on the amplitude of the outer cell of the meridional circulation, which depends on the compactness of the star as $U_r\propto \left[ 1-\frac{\Omega^2}{2\pi G \overline{\rho}}\right]$. For the same metallicities, the amplitude of the outer cell will be in a ratio 100~:~17~:~4~:~1 (Fig.~\ref{frurv}, \emph{centre}). This means that at $Z=0$, the core-envelope coupling is almost null. Figure~\ref{frurv} (\emph{right}) shows the evolution of the equatorial velocity during the MS for our $20\ M_{\odot}$ models. At standard metallicity, the meridional circulation is strong, but the mass loss is also strong, so the model spins down. At lower but non-zero metallicities, the mass loss drops, and though the core-envelope coupling is weak, the models spin up. At $Z=0$, the mass loss is null, but the model evolves almost in a regime of local conservation of angular momentum. Thus, the natural inflation of the radius during the MS leads to a spin down of the model. The model may reach the critical limit, but later in its MS evolution (for a given initial mass and velocity), and the mechanical mass loss remains very small.

\begin{figure}[b]
\begin{center}
 \includegraphics[width=.43\textwidth]{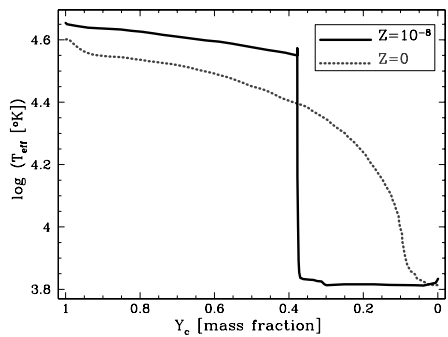}
 \caption{Evolution of $T_\mathrm{eff}$ during core He-burning for two models of $60\ M_{\odot}$. The $Z=10^{-8}$ model is taken from \citet{Hi07}.}
   \label{fteffy}
\end{center}
\end{figure}

At central H exhaustion, the core of the $Z=0$ model is already hot enough to burn some helium, so the transition between H-burning and He-burning is smooth, without any structural readjustment. The model remains in the blue part of the Hertzsprung-Russell diagram until very late in the central He-burning phase (Fig.~\ref{fteffy}). When it eventually evolves towards the red part, the outer convective zone is very thin, so the enrichment of the surface remains very low. Whereas at $Z=10^{-8}$ we get a strong post-MS mass loss, at $Z=0$ it is negligible.

Rotation alone seems thus to fail in providing a way to lose sufficient mass at $Z=0$.

\subsection{Two natural effects of rotation}

In the previous calculations, we have neglected two mechanisms that arise naturally with rotation and could change this picture.

The first one is the wind anisotropy \citep[see][]{Maeder99}: when a star rotates fast, it becomes oblate. As shown by von Zeipel theorem, the poles become hotter than the equator, and the radiative flux is no more spherically symmetric: it gets much stronger in the polar direction than in the equatorial plane. Since the mass is lost preferentially near the poles, it removes much less angular momentum than in the spherical configuration, and thus the star may reach the critical limit much earlier.

The second mechanism is the magnetic fields. According to \citet{Spruit02} the differential rotation may amplify an existing magnetic field through the Tayler-Spruit dynamo mechanism, and provide a strong core-envelope coupling. This coupling will be able to accelerate the surface very early in the evolution, and maintain the star at critical limit throughout its entire evolution.

\section{Evolution with anisotropical winds and magnetic fields}

\subsection{Ingredients}
With the help of the two effects mentioned above, we have computed an exploratory model. Its mass has been chosen to be $150\ M_{\odot}$ and the initial ratio between the equatorial velocity and the critical one to be $\upsilon_{\rm ini}/\upsilon_{\rm crit} = 0.56$. Let us mention that this ratio is higher than the one needed to account for the observed average velocities at solar metallicity, but it is still a ``reasonable'' value, not an extreme one.

The computation has been accomplished using the Geneva code with up to date nuclear reaction rates obtained with NETGEN (\url{http://www-astro.ulb.ac.be/Netgen/}). The opacity tables come from OPAL (\url{http://www-phys.llnl.gov/Research/OPAL/opal.html}) with the extension at low temperature by \citet{Ferg05}. The initial composition is $X=0.753$, $Y=0.247$ and of course $Z=0$.

The radiative mass loss prescription is an important ingredient of the modelization of massive stars. Here we have used \citet{Kudr02}. Since this prescription is not aimed at the case $Z=0$ strictly, we have used the same adaptations as in \citet{MarigoCK03}. The Wolf-Rayet (WR) mass loss rate is taken from \citet{NugLam00} with the metallicity scaling from \citet{EV06}. For the calculation, we have taken the effective surface metallicity $Z_{\rm eff}=(1-X-Y)_{\rm surf}$ so that the enrichment of the surface is accounted for. We must stress that this $Z_{\rm eff}$ is mainly composed by CNO elements but no iron. It is usually considered that WR winds are triggered by Fe lines, whereas the CNO lines determine only $\upsilon_{\infty}$, so we expect that no WR winds can take place at $Z=0$. But \citet{VdK05} have shown that when the metallicity gets really low, the CNO lines take over the role of Fe lines in the line driving, and this is what we have assumed here. 

When the model reaches the critical limit, the mechanical mass loss has been treated as described in \citet{MEM06}.

The treatment of anisotropic winds has been implemented as in \citet{Maeder02} and the effect of magnetic fields as in \citet{MM05rotBIII}.

\subsection{Result}

\begin{figure}[b]
\begin{center}
 \includegraphics[width=.75\textwidth]{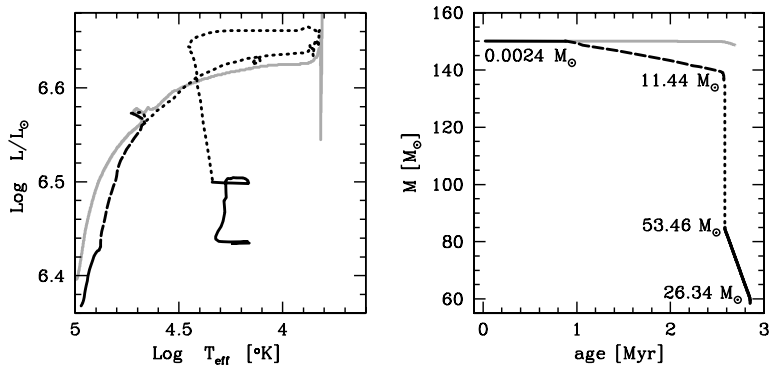}
 \caption{Black line: rotating model; \emph{continuous part}: beginning of MS ($X_{\rm c}=0.753$ down to 0.58; \emph{dashed part}: rest of the MS; \emph{dotted part}: beginning of core He-burning phase ($Y_{\rm c}=1.00$ down to 0.96); \emph{continuous part}: rest of the He-burning. Grey line: non-rotating model for comparison. \textbf{Left panel}: evolution in the Hertzsprung-Russell diagram; \textbf{Right panel}: evolution of the mass of the model. The mass indicated is the mass lost at each stage, not a summation.}
   \label{fevol150}
\end{center}
\end{figure}

In Fig. \ref{fevol150}, we present the evolution in the HR diagram (left panel) and the evolution of mass with time (right panel). The grey line shows a non-rotating model computed with the same physics for comparison.

During its whole evolution up to the end of core He-burning, the non-rotating model loses only $1.37\ M_{\odot}$. This illustrates the weakness of radiative winds at $Z=0$.

The evolution of the rotating model (black line) can be described by four distinct stages:
\begin{enumerate}
\item \emph{(continuous part}, lower left corner) The model starts its evolution on the MS with only radiative winds, losing only a little more than $0.002\ M_{\odot}$. During this stage, it accelerates quickly, mainly because of the strong coupling exerted by the magnetic fields.

\item \emph{(dashed part)} When its central content of hydrogen is still 0.58 in mass fraction, it reaches the critical velocity and starts losing mass by mechanical wind. It remains at the critical limit through the whole MS, but the mechanical wind removes only the most superficial layers that have become unbound, and less than 10\% of the initial mass is lost at that stage ($11.44\ M_{\odot}$). The model becomes also extremely luminous, and reaches the Eddington limit when 10\% of hydrogen remains in the core. Precisely, it is the so-called $\Omega\Gamma$-limit that is reached here. Due to the fast rotation, the maximum Eddington factor allowed is reduced; at the same time, because of the high luminosity, the critical velocity is reduced in comparison with the one derived from the classical $\Omega$-limit \citep[see][for details]{MM00}.

\item \emph{(dotted part)} The combustion of helium begins as soon as the hydrogen is exhausted in the core, then the radiative H-burning shell undergoes a CNO flash, setting the model on its redward journey. The model remains at the $\Omega\Gamma$-limit and loses a huge amount of mass. The strong magnetic coupling keeps bringing angular momentum to the surface and even the heavy mass loss is not able to let the model evolve away from the critical limit. The mass lost during that stage amounts to $53.46\ M_{\odot}$. When the model starts a blue hook in the HR diagram, its surface conditions become those of a WR star ($X_{surf} < 0.4$ and $T_{\rm eff} > 10'000$ K). The luminosity drops and takes the model away from the $\Gamma$-limit, marking the end of that stage.

\item \emph{(continuous part)} The rest of the core He-burning is spent in the WR conditions. The mass loss is strong but less than in the previous stage: another $26.34\ M_{\odot}$ are lost.
\end{enumerate}

At the end of core He-burning, the final mass of the model is only $M_{\rm fin}=58\ M_{\odot}$, already below the minimum $M_{\alpha}$ needed for PISN ($M_{\alpha} \geq 64\ M_{\odot}$). Note that the contraction of the core after helium exhaustion brings the model back to critical velocity, so this value for $M_{\rm fin}$ must be considered as an upper limit.
\section{Summary}

The model we presented here is exploratory. We cannot draw general conclusions from it. But our model shows that heavy mass loss is possible even at $Z=0$, and the answer to our title's question is: \emph{yes, under certain conditions, very massive stars can indeed avoid PISN}.

Some aspects need yet to be clarified. Is the WR mass loss rate we have used really valid? Can the CNO lines alone really drive a WR wind? In a more general perspective, we still lack a good mass loss recipe for the strict $Z=0$ case. A word of caution must also be cast on the inclusion of magnetic fields. The validity of the Tayler-Spruit dynamo is still under debate, and more work need to be done before we may confidently rely on results that have been obtained with the actual treatment. Moreover, there is not a clear consensus whether magnetic fields were present in the early Universe or not. In the Taylor-Spruit dynamo, the mechanism amplifies a pre-existing field, so if none were present at the start, we cannot use it.

Anyway, the physics used in the present model is today's ``state of the art'' and it is interesting to study what can be achieved with it. Our result is encouraging, because the computation has been accomplished with reasonable assumptions: the initial rotation rate used here was fast but not extreme, and the mechanisms called upon (anisotropy of the winds and magnetic fields) are not exotic ones, but two natural effects which arise when one treats properly the case of rotation.

After this first step, more work is needed. First, we have to refine the mass loss treatment at $\Omega\Gamma$-limit and check if the mass loss stays as strong as here. If it does so, we have to check if our result are valid in the whole PISN mass domain. Higher mass models should experience higher mass loss, but it is necessary to check if it would still be sufficient to help them avoid pair-instabilities.
\bibliographystyle{cup}
\bibliography{ekstrom}

\end{document}